\documentclass[12pt]{article}

\usepackage{graphicx}
\usepackage{epstopdf}
\usepackage[body={17.5cm, 21cm},right=2cm]{geometry}
\usepackage{amssymb}
\usepackage{amsmath}
\usepackage{psfrag}
\usepackage{epsfig}
\usepackage{cancel}
\usepackage{epstopdf}

\usepackage[all]{xy}

\newcommand{\bra}{\langle}
\newcommand{\ket}{\rangle}
\newcommand{\cH}{{\cal H}}
\newcommand{\cd}{{c^\dagger}}

\newcommand{\al}{\alpha}
\newcommand{\be}{\beta}
\newcommand{\ald}{\alpha^{\dagger}}
\newcommand{\bed}{\beta^{\dagger}}
\newcommand{\gad}{\gamma^{\dagger}}

\newcommand{\x}{{\bf x}}

\newcommand{\kk}{{\bf k}}
\newcommand{\ad}{a^\dagger}

\newcommand{\mb}{\overline{M}}

\newcommand{\bd}{b^\dagger}

\begin{document}
\setcounter{page}{0}
\thispagestyle{empty}

\begin{titlepage}

\begin{center}

\vspace{1.cm}

{\LARGE \sc{
Modeling a Particle Detector\\[2mm] in Field Theory
}}\\[1cm]
{\large Fabio Costa$^{\rm a, b}$ and Federico Piazza$^{\rm c}$}
\\[0.6cm]

\vspace{.2cm}
{\small \textit{$^{\rm a}$
Institute for Quantum Optics and Quantum Information,
Austrian Academy of Sciences, Boltzmanngasse 3, A-1090 Vienna,
Austria}}

{\small \textit{$^{\rm b}$
Faculty of Physics, University of Vienna,
Boltzmanngasse 5, A-1090 Vienna, Austria}}

{\small \textit{$^{\rm c}$
Perimeter Institute for Theoretical Physics
Waterloo, Ontario, N2L 2Y5, Canada}}

\end{center}

\vspace{.8cm}
\begin{abstract}
Particle detector models allow to give an operational definition to the particle content of a given quantum state of a field theory. 
The commonly adopted Unruh-DeWitt type of detector is known to undergo temporary transitions to excited states even when at rest and in the Minkowski vacuum. We argue that real detectors do not feature this property, as the configuration ``detector in its ground state + vacuum of the field" is generally a stable bound state of the underlying fundamental theory (e.g. the ground state-hydrogen atom in a suitable QED with electrons and protons) in the non-accelerated case. 
As a concrete example, we study a local relativistic field theory where a stable particle can capture a light quantum and form a quasi-stable state. As expected, to such a stable particle correspond energy eigenstates of the full theory, as is shown explicitly by using a dressed particle formalism at first order in perturbation theory.  We derive an effective model of detector (at rest) where the stable particle and the quasi-stable configurations correspond to the two internal levels, ``ground" and ``excited", of the detector.
\end{abstract}
 
\end{titlepage}

\section{Introduction}
 In Minkowski spacetime, particles can be identified with the excited modes of the field Hamiltonian. However, when field quantization is applied to general backgrounds, a univocal definition of particle is no longer possible \cite{birrell}. Still, with sound operational attitude, one can model a particle detector, calculate its response along some trajectory and associate a particle content to the corresponding observer/detector. A detector model is generally considered to be reasonable as long as no particles are revealed in the vacuum when the detector is at rest and \textit{long enough} times are considered; however, typical predictions include the possibility that, for short times, particles can be detected by a detector in inertial motion. Is this effect featured by real-world measuring devices, or is it only the result of assumptions specific to particular models? 

In this note we attempt to clarify some aspects of particle detector modeling and, in particular, emphasize the role of the eigenstates of the full Hamiltonian when the configurations of ``field+detector" are taken into consideration. In particular, we will deal with two levels of description: the first level is our \textit{bona fide} ``fundamental" theory, which we take as a weakly coupled QFT with three neutral scalar fields.
This theory features two stable particles (``$A$" and ``$C$") and one meta-stable state (``$B$"). The massive particle $A$ can capture the light quantum $C$ and form the unstable particle $B$. Stable particles correspond to eigenstates of the full Hamiltonian, as is shown explicitly by using a dressed particle formalism. Within this first level of description, we interpret the capture process as ``detection" of the particle $C$.
The second level of description is effective and features only a two-level particle detector and the (otherwise-) free field to be detected. The detector-field interaction is such that the transition rates $A \rightarrow B$ of the fundamental theory are faithfully reproduced, at the effective level,  as internal between the two levels (``ground state"--``excited") of the detector model. In this paper we deal only with inertial detectors in Minkowski spacetime.

The main observation at the basis of the present analysis is that the simple configuration ``the detector is in its ground state and the field is in the vacuum"  is stable and therefore should correspond to an \textit{eigenstate} of the Hamiltonian, both at the level of the detector model and at the level of the more fundamental interacting field theory. The commonly adopted Unruh-DeWitt detector~\cite{Unruh,dewitt} satisfies only a weaker version of this very basic request, namely, it sees no particles in the (Minkowski) vacuum after infinite time; however, for finite times, it always undergoes temporary transitions to excited states with finite, albeit small, probability. Curiously, the property that we are asking for is instead shared by the very first  photodetector model, proposed by Glauber in 1963 \cite{glauber}. After the works of Unruh and DeWitt, the Glauber detector -- in which the terms responsible for photon generation from the vacuum are absent --  has been considered as an approximation (the ``rotating wave approximation") of a more realistic detector.

Here we show that the Glauber type of the detector corresponds, at the deeper QFT-level description, to dealing with the properly ``dressed" states of the full theory. Therefore, we argue that the Glauber type of detector is more accurate for describing finite-time processes. The proposed detector model, although derived from a perfectly local relativistic field theory, does not couple to the local degrees of freedom of the field to be detected and, therefore, it is not {\it localized}~\cite{papero,papero2} in the usual sense. This is due to the fact that we are capturing and modeling the finite time behavior of the dressed -- as opposed to the bare -- states of the field theory itself.

The paper is organized as follows: in Sec. \ref{sec1} we analyze the Unruh-DeWitt model, showing how it fails the condition of not detecting particles in the vacuum. In the Sec. \ref{sec2} we introduce the toy field theory which models the detector at the fundamental level; we apply to such a theory a dressed particle formalism which allows to describe, at some given order in perturbation theory, the eigenstates/stable particles of the Hamiltonian. In Sec. \ref{sec3} we build the effective model where   the trasition rates of the underlying field theory are reproduced. Finally, in the last section, we discuss some of the implications of our analysis.

\section{Detector Models: a Critique} \label{sec1}

A model detector \cite{Unruh, dewitt,glauber} is a quantum system whose states live in a product Hilbert space $\cH_D \otimes \cH_\phi$ (\textit{i.e.} detector and field) and provided with a Hamiltonian operator $H_{m} = H^D_m + H^\phi_m + H^I_m$ (suffix $m$ stands for ``model''). In the simplest scenario, $\phi$ is an -- otherwise free -- scalar field, 
\begin{equation}
H^\phi_m = \int d^3k E(k) \cd_{\kk} c_{\kk}, \quad {\rm where} \quad E(k) = \sqrt{k^2 + m^2}
\end{equation} 
and $\cd_{\kk}$ and $c_{\kk}$ are the usual creation and annihilation operators,
\begin{displaymath}
	\phi(\x)= \frac{1}{(2\pi)^3}\int\frac{d^3k}{\sqrt{2 E(k)}}\left(\cd_{\kk}e^{i\kk\cdot\x}+c_{\kk}e^{-i\kk\cdot\x} \right)\;.
\end{displaymath}

The detector Hamiltonian $H^D_m$  accounts for  at least two energy levels: unexcited, $|0\ket_D$, and excited, $|E\ket_D$; say that $H_m^D |E\ket_D = \Delta E |E\ket_D$, $H_m^D |0\ket_D =0$. 
Regardless of the choice of $H^I_m$, the model state  $|0\ket_D \otimes |0\ket$ is thus interpreted, by construction, as ``the detector is in its ground state and the field is in its vacuum state".

The traditionally used Unruh--DeWitt detector features an interaction Hamiltonian of the type
\begin{equation} \label{unruh}
H^I_m \ =\ \sigma\, \phi(\x(t),t),
\end{equation}
where $\sigma$ is a self adjoint operator acting on $\cH_D$ and 
containing off diagonal elements and $\x(t)$ is the detector's trajectory. Without loss of generality, we can take $\sigma = \sigma_{\uparrow} + \sigma_{\downarrow}$, where $\sigma_{\uparrow} = |E\ket_D \bra0|$ and $\sigma_{\downarrow} = (\sigma_{\uparrow})^\dagger$.
The Hamiltonian \eqref{unruh} is based on the following requests:

\begin{enumerate}
	\item The detector is a quantum system with discrete energy levels.
	\item Transitions between different levels must be possible in order to account for particle absorption and emission (the simplified version containing two levels only allows single-particle detection).
	\item The detector interacts locally with the field.
	\item \textit{Asimptotically}, no transition has to take place when the detector is at rest.
\end{enumerate}

A striking feature of this traditionally adopted detector  
is that the state $|0\ket_D\otimes |0\ket $ is not an eigenstate of \eqref{unruh}, due to the presence of the creation operators $\cd_{\kk}$ inside $\phi$. Accordingly,
if the system is  initially prepared in the configuration 
$|0\ket_D \otimes |0\ket$, there is always a non vanishing transition rate to a state of type
$|E\ket_D \otimes |{\rm one\ particle}\ket$ at finite times, 
regardless of the state of motion of the detector. 
In the interaction picture, and at first order in perturbation theory,
the amplitude for this process reads
\begin{equation} \label{amp}
{\cal A}_\kk(2t)\ = \ -i \int_{-t}^t dt' \bra \kk|\otimes _D\!\bra E|\, H^I_m(t')\, |0\ket_D \otimes |0\ket ,
\end{equation}
where $ |\kk\ket = \cd_{\kk}|0\ket$, $H^I_m(t') = e^{i(H^D_m \tau(t) + H^\phi_m t)} H^I_m e^{-i(H^D_m \tau(t) + H^\phi_m t)}$ 
and $\tau$ is the proper time along the trajectory considered. If the detector is at rest,
${\cal A}_\kk(2t) \propto \sin[(\Delta E + E(k))t]/(\Delta E + E(k))$; only for $t\rightarrow\infty$, \textit{i.e.} $t\gg 1/\Delta E$, does ${\cal A}$ become proportional to a delta function of the positive quantity $\Delta E + E(k)$, and therefore vanishes. 
On the opposite, $|{\cal A}|$ keeps staying above zero
along accelerated trajectories, which is one of the several derivations of
the Unruh effect \cite{Unruh}.
Finite-time transitions and the 
Unruh response itself seem therefore to be related; they are both
consequences of the non-vanishing off diagonal elements $\bra\kk|\phi|0\ket$ of the interaction Hamiltonian \eqref{unruh}. 

In this paper we consider in detail the case of a detector at rest.
The possibility that a detector may ``click'' at finite times in the vacuum, but then ``erase'' the record later, looks rather 
misterious. One may object that the measurement process cannot be considered finished as long as the detector undergoes quantum fluctuations, so that only if $|{\cal A}_\kk(t)|$ stays definitely above zero in the future can we fairly say that a particle has been detected. However, in real apparatuses quantum coherence is usually destroyed over very short times by some amplification process; for decoherence times
$t_d \ll 1/\Delta E$, the vacuum fluctuations \eqref{amp} would become detectable and give rise to observable effects (according to \cite{rabi}, forthcoming experiments in circuit QED should be able to reveal such effects). 

It is also interesting to note that conjectured experimental proofs of vacuum entanglement rely on such ``vacuum dark counts''. For example, an experiment of entanglement distillation from the vacuum has been designed \cite{reznik} by using two Unruh-DeWitt detectors. The latter are kept at rest in Minkowski vacuum at a given relative distance and are both switched on and off at space-like separated events and for time intervals $\Delta t \ll 1/\Delta E$. Through the interaction with the vacuum the detectors get entangled between each other and the above described dark counts should show EPR-like correlations. If ever realized, this experiment would give operational meaning to the entanglement of the vacuum \cite{bombelli} between different regions of space. Here we are going to argue that realistic detectors have a more trivial response on the vacuum. As a corollary, 
our analysis suggests that the entanglement of the vacuum has no operational implications and cannot possibly be used to create testable correlations.

An elementary and reasonable detector one may think of is a hydrogen atom that, 
by absorbing a photon, can make a transition to an excited state. We can think of a consistent
 QED theory with two Dirac fields of opposite charges (electrons and protons) and appropriate 
masses; the hydrogen atom in its ground state is arguably
 contemplated in the spectrum of that theory as a stable bound state. 
When written in terms of fundamental fields, we therefore expect
the model state $|0\ket_D \otimes |0\ket$ to concretely correspond to
a stable state, \textit{i.e.}, strictly, 
an eigenstate, of the full Hamiltonian: this is what the unexcited hydrogen atom is in QED and the detector is not in the Unruh-DeWitt model. It is plausible that even more realistic detectors, such as a block of germanium crystal, correspond 
to stable bound states in appropriate QED-like theories.

The amplitude \eqref{unruh} is clearly analogous to the usual perturbative calculation of $S$ matrix elements; in that formalism, under the consistent assumption of adiabatic switching of the interactions, asymptotic in- and out- states are borrowed from the free theory.  Take a $\lambda \phi^4$ theory as an example. Similarly to the Unruh-DeWitt detector, the Hamiltonian features off diagonal elements, among others, of the type $\bra {\rm four\ particles}|H |0\ket$. The latter, however, are just the matrix elements between the unphysical states of the free theory: we know that
$\lambda \phi^4$ has a stable vacuum and stable single-particle states; when written in terms of those, the full Hamiltonian has, by definition, only diagonal elements. 
 One would come to wrong conclusions if the states of the free theory were used to study finite-time processes.

The above reasoning brings us to postulate, as necessary for a good  model detector, that the configuration ``unclicked detector + vacuum",  $|0\ket_D \otimes |0\ket$, 
be an eigenstate of the Hamiltonian $H_m$ of the model at rest. We will refer to that as \textit{the Frog Principle}\footnote{Some frogs are known to have eyes sensitive enough to detect a single photon~\cite{frog}. The postulated triviality of a good detector's response on the vacuum -- and the absence of dark counts -- may be pictorially rephrased as ``a frog does not see photons when it's dark".}. This principle can be regarded as a stronger version of the above Condition 4: the condition of not clicking in the vacuum is extended from infinite to finite times. It seems that if we make Condition 4 stronger, all four requirements cannot be satisfied simultaneously. What seems problematic, in particular, is to reconcile locality (Condition 3) with the frog principle. It is indeed a well known consequence of the Reeh-Schlieder theorem that no nontrivial, positive, local observable can exist which has zero expectation value on the vacuum~\cite{reeh, redhead}. This implies that, if a detector performs a von Neumann-type measurement corresponding to a projector $\Pi$, then, either $\bra 0|\Pi|0\ket \neq 0 $ or $\Pi$ is not local (see also Ref.~\cite{terno}, Sec. IV E). We suggest that, for detector models, Condition 3 should indeed be dropped, and that this is perfectly compatible with local relativistic quantum field theory.
In order to make our point stronger, in the following we consider a toy -- local relativistic --``fundamental" field theory where a stable particle plays the role of the detector in its ground state and the 
detection process corresponds to the capture of a light particle and the formation of a meta-stable state. 
 We will then provide a two-level effective detector model faithfully reproducing the 
detection rates of the fundamental theory and satisfying the \textit{Frog Principle}. A similar toy field theory was sketched already in Unruh's celebrated paper \cite{Unruh}. The crucial difference here is that we use ``dressed" rather than ``bare" states since we aim to capture and model the response of a physical particle.

Studies of Unruh-DeWitt models have gone beyond the perturbative amplitude \eqref{amp}, exact and numerical solutions are available for a variety of trajectories (see e.g. \cite{hu}). Considering exact asymptotic states is particularly appropriate since, in most cases, preparing the system into the state $|0\ket_D\otimes|0\ket$ and then  switching on the interaction is not realistically possible. Crucially, the ``dressed" stable configurations of the Unruh-DeWitt detector depend on the trajectory considered and exhibit radiation at infinity in the accelerated case.  In this paper we show that if dressing is consistently done -- in the first place -- on 
all sectors of the underlying field theory, this produces a different detector model altogether, \textit{i.e.} a different model Hamiltonian $H_m^I$  (eq. \ref{ours} below).

\section{The Toy Field Theory} \label{sec2}

Beside the already introduced light field to be detected, $\phi(x)$ of mass $m$ (sector ``$C$'' of the theory), we introduce two other neutral scalars, $\chi(x)$, of mass $M$ (sector ``$A$''), and 
$\eta(x)$ of mass $\mb$ (sector ``$B$''). We choose a local interaction  of the type ${\cal L}^I(x) \sim -\mu\, \chi(x)\eta(x)\phi(x)$, the coupling $\mu$ being small with respect to the other masses\footnote{Strictly speaking, this potential is not bounded from below \textit{e.g.} along the direction $\chi = \phi$, $\eta = -\phi$. 
However, the tunneling decay rate of the perturbative vacuum is suppressed by an exponential factor of at least $e^{-m^2/\mu^2}$ which we fix to be
small enough to be irrelevant. Moreover, we can always stabilize the potential with higher order terms that will not be relevant for present purposes.}. 
The full Hamiltonian reads $H = H^0+ H^I$, where $H^0 = H^A + H^B + H^C$, and
\begin{align}\label{1}
H^A =& \int d^3 k\, w(k) \ad_{\kk} a_{\kk}, \quad w(k)^2 = k^2 + M^2,\\
H^B =& \int d^3 k\, W(k) \bd_{\kk} b_{\kk}, \quad W(k)^2 = k^2 + \mb^2,\\
H^C =& \int d^3 k\, E(k) \cd_{\kk} c_{\kk}, \quad E(k)^2 = k^2 + m^2,
\end{align}
\begin{multline} \label{hi}
\quad H^I = \, (2 \pi)^{3/2} \mu \int d^3x\, \phi(\x) \eta(\x) \chi(\x) \\[2mm]
= \mu \int\frac{d[k_1k_2k_3]}{\sqrt{2 w(k_1)\, 2 W(k_2)\, 2 E(k_3)}}\, 
 (a_{\kk_1}+\ad_{-\kk_1}) (b_{\kk_2}+\bd_{-\kk_2}) (c_{\kk_3}+\cd_{-\kk_3}).\quad
\end{multline} In the above expression $d[k_1k_2k_3] = d^3 k_1\, d^3 k_2\, d^3 k_3\, \delta^3 (\kk_1 + \kk_2 +\kk_3)$ is the volume element on the momentum shell.
Creation and annihilations operators have been introduced in the usual way and satisfy usual commutation relations. 
In the picture we have in mind $\phi$ is a light field ($m\ll M,\mb$) that can be captured by an $A$-particle and form a $B$ meta-stable state. The mass difference $\Delta M = \mb - M$ is therefore supposed to be of the same order as -- but slightly bigger than -- $m$, $\mb > M + m$. In order to allow a perturbative treatement, the coupling $\mu$ is taken much smaller than the other masses, $\mu\ll m$. Particles $A$ and $C$ are stable. $C$ cannot decay to anything else for kinematical reasons. Moreover, processes such as $A\rightarrow 2C$ are not allowed by the form of the interaction:
formally, the discrete symmetry $\phi \rightarrow -\phi$, 
$\chi \rightarrow -\chi$, $\eta \rightarrow \eta$ $+\, permutations$ is protected. 

We aim to give an effective description of the first order in $\mu$- $ABC$ dynamics in which sectors $A$ and $B$ are described as ``internal" to the model detector and in such a way that the transition amplitudes are faithfully reproduced. The one-particle sector $A$ is the detector in its ground state. The excited detector is described instead by the meta-stable configurations of the $B$ sector. With the above assumed relations among the mass parameters, the decay rate of a $B$-particle is $\Gamma_{B} \sim \mu^2 \Delta M/M^2$. At the expense of detector's efficiency, we can assume $B$'s lifetime $\tau_B \sim 1/\Gamma_{B}$ to be long enough
for the detector to be considered as ``permanently clicked'' for all practical purposes. 

As announced, we want the one particle- $A$ sector of this theory to correspond to the state 
$|0\ket_D \otimes |0\ket$ of the model detector. However, $\ad_{\kk} |\Omega\ket$ (here $|\Omega\ket$ is the field theory vacuum, as opposed to the vacuum $|0\ket$ of the field $\phi$ in the detector model \eqref{unruh}) is an eigenstate of the free theory but not of $H^I$,
due to the presence in \eqref{hi}, e.g.,  of terms such as $a \bd \cd$.  On the other hand, we know that the $A$-particle is stable and therefore corresponds to a set of eigenstates also in the full theory. Such states can be expressed, order by order in perturbation theory, through a ``clothing'' or ``dressing'' transformation\footnote{A QFT formulation in terms of clothed particles dates back to the late 50s \cite{green}, although similar approaches date even earlier. The beautiful paper \cite{wilson} explores a similar 
transformation at the pure level of matrix elements. Our two main references are \cite{shebe,stefa}, where a complete bibliography on the subject can be found.}.
For this purpose, we act with a unitary transformation $U$ on the whole Hilbert space,
$|\Omega\ket\rightarrow|\Omega_d\ket = U |\Omega\ket,\, \ad \rightarrow \alpha^\dagger = U \ad U^\dagger, \bd \rightarrow \beta^\dagger = U \bd U^\dagger , \cd \rightarrow \gamma^\dagger = U \cd U^\dagger\, {\rm etc}\dots $
and impose that the ``dressed'' states $|\Omega_d\ket, \alpha^\dagger|\Omega_d\ket, \gamma^\dagger|\Omega_d\ket$ be eigenstates of the full Hamiltonian. On the other hand, $\beta^\dagger_d|\Omega_d\ket$ won't be an eigenstate cause $B$-particles are unstable. Following \cite{shebe}, 
we write $U = e^R$, where $R$ is an anti-hermitian operator,
$R = {\cal R} - {\cal R}^\dagger$ that can be written at first order in $\mu$ in terms of the bare operators. We make the ansatz
\begin{equation} \label{r}
{\cal R} = \mu \int d[k_1k_2k_3]\left(F_1 a_{\kk_1} b_{\kk_2} c_{\kk_3} + F_2 a_{\kk_1} b_{\kk_2} \cd_{-\kk_3} \right. \\ \left.+ F_3 a_{\kk_1} \bd_{-\kk_2} c_{\kk_3} + F_4 \ad_{-\kk_1} b_{\kk_2} c_{\kk_3}\right),
\end{equation}
where the $F$s are functions of the moduli $k_1$, $k_2$ and $k_3$, regular on the momentum shell $\kk_1 + \kk_2 +\kk_3 =0$.
Instead of transforming the states we can equivalently transform the Hamiltonian although, in spirit, what we are doing is really rewriting the same Hamiltonian in terms of the dressed operators $\alpha$, $\beta$, etc\dots To first order in $\mu$ the transformation reads $H\rightarrow H_d = H + [R, H]$. The zeroth order free part $H^0$ is left unchanged by this transformation. The interaction part gets a contribution of the type $H^I \rightarrow H^I_d = H^I + [R, H^0]$. To give an example, 
let's see this commutator in detail for terms of type $\ad bc$ and $a \bd \cd$ inside $R$. Terms of this type would make the $A$-particle decay into $B+C$ and therefore are not physical. 
\begin{equation} \label{trans}
[R_4, H^0] = \mu \int d[k_1k_2k_3] \left(- w(k_1) + W(k_2) + E(k_3)\right)
 \left(\ad_{-\kk_1} b_{\kk_2} c_{\kk_3} + a_{\kk_1} \bd_{-\kk_2} \cd_{-\kk_3}\right)F_4 \; .
\end{equation}
Note that, by setting $1/F_4 =  (-w(k_1) + W(k_2) + E(k_3))\sqrt{2 w(k_1)\, 2 W(k_2)\, 2 E(k_3)}$ in \eqref{r}, we can get rid of the corresponding terms inside $H^I$.

Other terms in $H^I$ get contributions similar to \eqref{trans}, except that the energies
$w$, $W$ and $E$ appear in different combinations \textit{i.e.} with appropriate relative signs.
Crucially, we cannot get rid of the term $a \bd c$, $\ad b \cd$, since the corresponding combination of energies, $w(k_1) - W(k_2) + E(k_3)$, vanishes on a subset of the momentum shell and the function $F_3$ would be singular there. 
Note that bare and dressed particles are bound to give the same S-matrix elements and decay rates, since the ``good terms" such as $a \bd c$, $\ad b \cd$ can only get harmless corrections that vanish on the energy shell! By setting $F_3=0$ in \eqref{r} we get the following dressed interaction Hamiltonian:
\begin{equation} \label{interaction}
H^I_d = \mu \int \frac{d[k_1k_2k_3]}{\sqrt{2 w(k_1)\, 2 W(k_2)\, 2 E(k_3)}}
\left(\al_{\kk_1} \bed_{-\kk_2} \gamma_{\kk_3} + \ald_{-\kk_1} \be_{\kk_2} \gad_{-\kk_3}\right)\, .
\end{equation}
The above operator is equal to the original Hamiltonian $H^I$ \eqref{hi} up to first order in $\mu$. A drawback of this formalism is that it gets rather involved at higher orders: new dressed operators and Hamiltonians have to be derived at each step. Lorentz invariance is guaranteed, since the dressing transformation $U$ preserves the commutation relations among the generators of the Poincar\'e group. However, as opposed to \eqref{hi}, \eqref{interaction} is not written in the local form $\int d^3 x V(\x)$, $V(x)$ being a scalar commuting at space-like separated events.
What is important here is that $H^I_d$ makes the stability of the  $A$ and $C$ sectors manifest and reproduces the dynamics with the required  accuracy.

\section{The Effective Detector Model} \label{sec3}

We are now ready to build our detector. We first specify the state of the theory that matches the state $|0\ket_D\otimes |0\ket$ of the detector model. In momentum space this will be expressed by
\begin{equation}
|0\ket_D \otimes |0\ket \ \simeq\ |g\ket_A \otimes |0\ket_B \otimes |0\ket_C =\int d^3k g(\kk) \ald_\kk |\Omega\ket \; .
	\label{init}
\end{equation}
It is not too restrictive to choose the detector at rest in a spherically symmetric configuration centered around some point in space $\x$, $i.e.$ $g(\kk)= g(k) e^{-i \kk \x}$, $g(k)$ being a real function.
As this state may well describe a macroscopic object, we can also assume the momentum fluctuations to be small compared to its mass (or, equivalently, the spatial extension to be much larger than the Compton wavelength). This is accomplished by a distribution $g(k)$ non vanishing only for $k^2 \ll M^2$, which makes the above state also an approximate eigenstate of the free evolution.

In order to study detector's response we now populate also the $C$-sector and consider the state $|\psi\ket = |g\ket_A \otimes |0\ket_B \otimes |f\ket_C$, where $|f\ket_C = \int d^3k f(\kk) \gad_\kk|0\ket_C$ and now $f$ can be centered around some $\kk \neq 0$. Still,
we take the energy of the particle to be detected much smaller than the mass of the detector, so that typically $f(\kk)$ is nonzero only for $E(k) \ll M$. 
In interaction picture the evolution of $|\psi\ket$ reads
$|\psi(2t)\ket = 
	( 1 -i \int^{t}_{-t}dt'H^I_d(t')) |\psi\ket$. The interaction picture Hamiltonian 
$H^I_d(t)$ is, in form, very similar to \eqref{interaction}, with the difference that the operators inside the brackets get a phase factor, \textit{i.e.}  $e^{i\Omega t}\al_{\kk_1}\bed_{-\kk_2}\gamma_{\kk_3}+ c.c.$, where 
$\Omega (k_1,k_2,k_3)= -w(k_1) + W(k_2)-E(k_3)$.
The amplitude ${\cal A}_\kk(2t)\equiv \bra \Omega_d| \be_\kk|\psi(2t)\ket$ for the creation 
of a $B$ particle of momentum $\kk$ thus reads
\begin{equation}
{\cal A}_\kk(2t)	=
 -2i\mu \int \frac{d^3k_c}{\sqrt{2 w(k_a) 2 W (k) 2 E(k_c)}} g(\kk_a)f(\kk_c)\frac{\sin \Omega(k_a, k, k_c)t}{\Omega(k_a, k, k_c)}.
 \label{final}
\end{equation}
In the above formula $\kk_a = \kk - \kk_c$.
Under the above assumptions, the functions $g$, $f$ cut the high momenta 
in the integral, so that we can make the following approximations: 
$w(k_a) \simeq M$, $W(k) \simeq \mb$, $\Omega \simeq \mb-M-E(k_c)=\Delta M - E(k_c) \equiv \Omega(k_c)$.

We now want to consider as ``detection" all possible final states of the $B$ field, regardless of the small recoils $\kk$ that the $A$-$B$ particle gets from the $C$ particle. When we integrate the squared amplitude 
\eqref{final} to get the detection probability  $P(2t)= \int d^3k |{\cal A}_\kk(2t)|^2$, there appears an interference term of the form $\int d^3k g^*(\kk-\kk_c)g(\kk-\kk'_c)$; this term cannot be reproduced by detector models where such recoil is just ignored. However, it looks reasonable to assume that $f$ be much less spread than $g$, since the spread in the momenta is naturally weighted by the respective masses.  Under this assumption, and recalling that $g(\kk) = g(k) e^{-i \kk \x}$, inside the expression for $P(2t)$ we always have $g(|\kk + \kk_c - \kk'_c |) \simeq g(k)$, and so we can put $\int d^3k g^*(\kk-\kk_c)g(\kk-\kk'_c) \simeq 
e^{i(\kk_c - \kk'_c) \x}$. In other words, the configuration $g(k)$ of the $A$ particle becomes irrelevant in the process whenever the light quantum has a much more definite momentum. Therefore, in the detector model that follows, the $\x$ variable is effectively coarse-grained by the typical spread $1/\Delta k_c$ of the particles that are detected. In the limit where $f(\kk_c) = \delta^3(\kk_c - \kk_{\rm particle})$ the $\x$ dependence drops from the rate and detector's position becomes irrelevant. 
The two integrals inside $P(2t)$ factorize and we finally obtain 
\begin{equation}
		P(2t)= \frac{\mu^2}{M^2}	\left|\int \frac{d^3k}{\sqrt{2E(\kk)}}f(\kk)e^{i\kk\cdot\x}\frac{\sin \,   \Omega(\kk)t}{\Omega(\kk)}\right|^2
		\label{prob}
\end{equation}

Our model detector has to reproduce the same detection rate for a generic initial state 
$|0\ket_D \otimes  |f\ket$, where $ |f\ket = \int d^3k f(\kk)  \gad_\kk |0\ket$ is the field state in the model. This is achieved through the effective interaction Hamiltonian
\begin{equation} \label{ours}
		H^I_m= \frac{\mu}{M}\left(\sigma_{\uparrow} \Phi^+(\x) + \sigma_{\downarrow} \Phi^-(\x)\right),
\end{equation}
where we recall that $\sigma_{\uparrow}$, $\sigma_{\downarrow}$ are the raising and lowering operators of the two level detector and the energy gap inside the detector is $\Delta E = \Delta M$. The complex fields $\Phi^+(\x) $  and $\Phi^-(\x) $ are defined in terms of the dressed annihilators as $\Phi^+(\x)=\int d^3k e^{i\kk \cdot \x}\gamma_\kk/\sqrt{2E(k)}$, $\Phi^- = (\Phi^+)^\dagger$. Eq. \eqref{ours} has the same matrix structure as \eqref{interaction}, where the $A\rightarrow B$ transition between Fock spaces is modeled inside a two level system through the raising operator $\sigma_{\uparrow}$.  

Dressing has effectively produced a different $ABC$- partition of the original local field theory \eqref{hi}; as a consequence, $\Phi^+(\x) $  and $\Phi^-(\x) $ are not the positive and negative energy part of the local field $\phi$, because they are built with the dressed operators $\gamma$ and $\gamma^\dagger$. However, at the pure level of the detector model, we should not care any more about the underlying theory, and we can still define an -- otherwise free -- field $\Phi(x) = \Phi^+(x) + \Phi^-(y)$  coupled to the detector through \eqref{ours}. Our local field theory \eqref{hi} has effectively produced a non-local detector for the field $\Phi$.

\section{Discussion}

In this paper we have built a detector model having the same response as a physical ``dressed" -- rather than ``bare" -- particle in an interacting field theory. 
More generally, we have tried to draw attention to the privileged role of the full theory's Hamiltonian eigenstates in describing typical objects and measuring devices. Of course, a quantum system can be considered in arbitrary states. However, what seems peculiar of field theory at low densities is that generic states generally evolve by radiating away decay products  until we are left, in a sufficiently large region of space, with an approximate eigenstate--field configuration. 
For simplicity, we have considered a weakly interacting toy-theory, although such arguments are known to apply even more dramatically to strongly interacting ones (scattering products ``hadronize" very rapidly if they are not QCD eigenstates).

Our detector model \eqref{ours} clearly obeys the \textit{Frog Principle}, as the state $|0\ket_D\otimes |0\ket$ is stable and no transitions can possibly occur at any finite time. This statement is valid at the level of the matrix elements of our effective Hamiltonian \eqref{ours} and, therefore, is independent of the state of motion of the detector. However, our derivation is fully consistent only in the case of an inertial detector, since this is the natural state of motion of the $A$-particle under the only influence of the field theory Hamiltonian. In order to study what happens under acceleration, one should consider, case by case, how this acceleration is consistently induced on the particle/detector. For example, one can make our $A$ and $B$ fields charged and make the detector accelerate under the effect of an external classical electric field. An analogous set-up has been considered in~\cite{francese}, where a direct relation between the Unruh and the Schwinger effects has been highlighted. It would be interesting, however, to study such an accelerating model not only in terms of asymptotic ``free" states, and consider also its short time behavior in terms of ``dressed" objects. 
The prediction that accelerated observers detect thermal radiation in the Minkowski vacuum is strong of many independent derivations. Its universality can be traced back to the fact that the response of a detector only depends on the characteristics of the Wightman function describing the field's correlations along the given trajectory and is therefore independent of the particular model, as long as the coupling detector-field is local~\cite{akhmedov}.
However, the most striking outcome of this paper is the non-local nature of the detector \eqref{ours}, despite the fundamental field theory \eqref{hi} from which it is derived is perfectly local and relativistic. 

More generally, our analysis seems to suggest that real measuring devices have no direct access to the local degrees of freedom $\phi(x)$ and effectively ``see" only the positive energy fields $\Phi^+$ of the dressed quanta.  Such a circumstance was already pointed out very clearly\footnote{ ``It has become customary, in discussions of classical theory, to regard the electric field ${\bf E}(\x, t)$ as the quantity one measures experimentally, and to think of the complex fields ${\bf E}^\pm(\x, t)$ as convenient, but fictitious, mathematical construcions. Such an attitude can only be held in the classical domain [\dots]. Where quantum phenomena are important the situation is usually quite different. [\dots] The use of any absorbtion process, such
as photoionization, means in effect that the field we are measuring is the one associated with photon annihilation, the complex field ${\bf E}^+(\x, t)$" \cite{glauber}.} by Glauber in his pioneering paper \cite{glauber}, where, in fact, a photodetector model analogous to \eqref{ours} is introduced.
In QFT it is customary to call ``observable"  the degrees of freedom associated with the local relativistic fields $\phi$. However,  in the constructive approach e.g. of the beautiful book \cite{wein},  the entire QFT formalism is built in order to give account for scattering experiments and decay processes, and the relativistic fields $\phi$ are introduced at some later stage (Chap. 5) with the purpose of writing Lorentz invariant interactions. The latter are in fact guaranteed if the Hamiltonian is local in the field $\phi$.  Here the local form of the Hamiltonian is not called into question and constitutes our starting point \eqref{hi}. What our analysis seems to question is whether  the local degrees of freedom are actually observable.

\section*{Acknowledgments}

We thank Dario Benedetti, {\v C}aslav Brukner, Sergio Cacciatori, Denis Comelli, Florian Conrady, Laurent Freidel, Maurizio Gasperini, Bei-Lok Hu, Ted Jacobson, Raymond Laflamme, Matthew Leifer, Xiao Liu, Christian Marinoni, Giovanni Marozzi, Matteo Nespoli, Michael Nielsen, Stefano Osnaghi, Tomasz Paterek, Maxim Pospolev, Carlo Rovelli, Alexandr Shebeko, Lee Smolin, Simone Speziale, Rafael Sorkin, Eugene Stefanovich, Gerard `t Hooft, Andrew Tolley, Enrico Trincherini, Bill Unruh, Gabriele Veneziano, Alberto Zaffaroni for useful conversations or correspondence. We are also grateful to Steven Weinberg for his inspiring textbook on quantum field theory \cite{wein}. Research at Perimeter Institute is supported by the Government of Canada through Industry Canada and by the Province of Ontario through the Ministry of Research and Innovation.

\end{document}